\documentclass{ifacconf}

\usepackage{graphicx} 
\usepackage{natbib}

\usepackage{amsmath} 
\usepackage[utf8]{inputenc} 
\usepackage{float}

\usepackage{enumitem}
\usepackage{xspace}
\def\url#1{\expandafter\string\csname #1\endcsname}

\begin{document}

\begin{frontmatter}

\title{FLOreS - Fractional order loop shaping MATLAB toolbox}

\author[First]{Lennart van Duist}
\author[Second]{Gijs van der Gugten}
\author[Third]{Daan Toten}
\author[Fourth]{Niranjan Saikumar}
\author[Fifth]{Hassan HosseinNia}

\address[First]{3ME, TU Delft (e-mail: info@duistech.com)}
\address[Second]{3ME, TU Delft (e-mail: G.vanderGugten@student.tudelft.nl)}
\address[Third]{3ME, TU Delft (e-mail: D.F.C.Toten@student.tudelft.nl)}
\address[Fourth]{PME, 3ME, TU Delft (e-mail: N.Saikumar@tudelft.nl)}
\address[Fifth]{PME, 3ME, TU Delft (e-mail: S.H.HosseinNiaKani@tudelft.nl)}

\begin{abstract}
\normalsize
{A novel toolbox named FLOreS is presented for intuitive design of fractional order controllers (FOC) using industry standard loop shaping technique. This will allow control engineers to use frequency response data (FRD) of the plant to design FOCs by shaping the open loop to meet the necessary specifications of stability, robustness, tracking, precision and bandwidth. FLOreS provides a graphical approach using closed-loop sensitivity functions for overall insight into system performance. The main advantage over existing optimization toolboxes for FOC is that the engineer can use prior knowledge and expertise of plant during design of FOC. Different approximation methods for fractional order filters are also included for greater freedom of final implementation. This combined with the included example plants enables additionally to be used as an educational tool. FLOreS has been used for design and implementation of both integer and fractional order controllers on a precision stage to prove industry readiness.
}

\end{abstract}


\end{frontmatter}

\section{Introduction}

Fractional order controllers (FOC) are gaining popularity in controls with active research and implementation in various applications. FOCs provide additional capability and flexibility in design and tuning compared to their integer-order counterparts (IOC). In fact, FOCs form the superset for IOCs and the former have found successful implementation in (\cite{delavari2012adaptive,hosseinnia2013fractional,tejado2011low,zamani2009design,hamamci2007algorithm,zhao2005fractional,saikumargeneralized,monje2010fractional,caponetto2010fractional}). Several works specifically target design, tuning and optimization of FOCs showcasing their growing importance (\cite{chen2009fractional,padula2011tuning,lee2010fractional}).

Although FOCs are gaining traction in academia, the same is not seen in the industry where integer order PID is used in nearly 95\% of applications. While some of the factors contributing to this are simplicity of design, robustness of PID; the main factor is that PID can be tuned using loop-shaping and tested tools exist which enable this. Loop-shaping is an intuitive method for designing and tuning controllers where the frequency response of plant is shaped in order to get the required closed loop properties (\cite{astrom2010feedback}). With loop-shaping, the control engineer uses experience in order to design the controllers with prior knowledge and expertise of the plant. Tested toolboxes such as ShapeIt (\cite{bruijnen2006optimization}) further aid the engineer in this process and allow for overall performance analysis.

Several toolboxes have been developed to aid design of FOCs. The popular ones include CRONE toolbox (\cite{oustaloup2000crone}), FOMCON (\cite{tepljakov2011fomcon}), NINTEGER (\cite{valerio2004ninteger}) and FOTF (\cite{moroz2017fotf}). However, these tools focus on designing FOCs from the time domain perspective which is not popular in the industry. Frequency domain based loop shaping tools only exist for IOCs. No existing toolbox combines FOCs with loop shaping which would enable transition of FOCs from academia to industry. This paper presents such a loop shaping toolbox --- FLOreS. Different approximation methods necessary for practical implementation of FOCs have also been included. FLOreS uses a graphical approach making it also useful as an educational tool. FLOreS is also used to design an FOC and IOC for a practical system and results from FLOreS are compared with practical results to show industry readiness. It must be noted that this paper does not focus on whether better performance can be achieved with FOCs compared to IOCs or on design techniques for FOCs. Instead, the focus is on filling the gap to enable industry engineers to design FOCs using loop-shaping method.

\section{preliminaries}

\subsection{Loop-Shaping}

Loop-shaping is an industry popular technique where the control engineer uses a graphical approach to shape the system open loop in order to meet the required specifications. These specifications include bandwidth; stability and robustness in terms of gain margin, phase margin and modulus margin; disturbance rejection and noise attenuation. These behaviours are represented graphically using Bode, Nyquist and Nichols plots.

While design is mainly carried out on open loop, plots of closed-loop sensitivity functions allow for overall performance analysis. The various sensitivity functions are as given below.

\begin{align}
\begin{split}\label{eq:1}
	\frac{y}{r} ={}\frac{PC}{1 + PC} \hspace{.2cm} \text{ complementary sensitivity function}
\end{split}\\
\begin{split}\label{eq:2}
	\frac{y}{d} = {}\frac{P}{1 + PC} \hspace{.2cm}  \text{ process sensitivity function}
\end{split}\\
\begin{split}\label{eq:3}
    \frac{u}{r} = {}\frac{C}{1 + PC} \hspace{.2cm} \text{ control sensitivity function} 
\end{split}\\
\begin{split}\label{eq:4}
    \frac{y}{n} = {}\frac{1}{1 + PC} \hspace{.2cm} \text{ sensitivity function}
\end{split}
\end{align}

In which the y, r, d and n are output, input, disturbance and noise respectively and P and C are plant and controller respectively.

\subsection{Fractional order controllers}

While the idea of non-integer orders for differentiation was raised as early as 1695 by Leibniz, this was only formalized with the works of Liuville and Riemann in the 19th century. The resulting fractional order calculus has been used in controls for both modelling of systems and design of FOCs. Fractional order calculus can be generalized in time domain as:

\begin{equation}
D^\alpha=
  \begin{cases}
    \frac{\text{d}^\alpha}{\text{d}t^\alpha}  & \quad \alpha>0,\\
    1  & \quad \alpha = 0,\\
    \int\limits_a^t(\text{d}\tau)^{-\alpha}  & \quad \alpha<0,\\
  \end{cases}
\end{equation}
\cite{caponetto2010fractional}

While the definition and interpretation of fractional order differentiation in time domain is useful, it is the laplace transform of fractional integral of a function which is useful for controller design using loop-shaping and this is given as
\begin{equation}
\mathcal{L}\{I^\alpha_0 f(t)\} = \frac{1}{s^\alpha} F(s)
\end{equation}

From the loop-shaping perspective, while an integrator has $-20\ dB/decade$ slope in magnitude and $-90^\circ$ phase, a fractional order integrator of order $\lambda$ has $-20\lambda\ dB/decade$ slope and $-90\lambda^\circ$ phase. This interpretation allows for effective adoption of fractional order calculus in design of controllers. As an example, the industry popular PID controller is given as:

\begin{equation}
IoPID = K_p \Bigg(1 + \frac{\omega_i}{s}\Bigg)\Bigg(\frac{1 + \frac{s}{\omega_d}}{1 + \frac{s}{\omega_t}}\Bigg)(other\ filters)
\end{equation}
where $\omega_i$ is integral action cut-off frequency, $\omega_d$ and $\omega_t$ are frequencies at which derivative action starts and is tamed respectively. Other filters including notch, anti-notch, low-pass filters may be used to satisfy other performance specifications. This representation of PID is used instead of the parallel representation because all the parameters of the equation directly relate to the frequency values of the filters used. From the Laplace transform of fractional calculus, PID can easily be extended to obtain Fractional order PID (FoPID) as:

\begin{equation}
FoPID = K_p \Bigg(1 + \frac{\omega_i}{s}\Bigg)^\lambda\Bigg(\frac{1 + \frac{s}{\omega_d}}{1 + \frac{s}{\omega_t}}\Bigg)^\alpha(other\ filters)
\end{equation}
resulting in non-integer integral and derivative actions. This provides two additional parameters during design and hence increases freedom and flexibility in shaping the open loop. It can be clearly seen that when $\alpha$ and $\lambda$ are 1, integer order PID is obtained.

\subsection{Approximation methods}

The Laplace interpretation of fractional order calculus allows for simple and easy visualization of fractional order integral and derivative actions. However, these cannot be directly implemented practically. Approximation techniques are required where fractional order transfer functions are approximated by integer order ones. While several approximation techniques exist, the most popular ones which are also included in FLOreS are explained here. The approximated transfer functions are of higher order and accuracy of approximation is dependent on the allowed order and this, in turn, depends on available computational capacity. Further, all approximation techniques work only within a range of frequencies.

Approximation techniques can be divided into continuous and discrete methods. Since most implementations in industry and academia today are carried out in discrete domain; this allows the engineer to either obtain the approximated transfer function in continuous domain and then convert it to discrete or directly obtain approximation in discrete domain. To provide this freedom to the engineer, both have been included in FLOreS.

\begin{description}[leftmargin=*]
\item[Continuous approximation methods:]

\item[Crone approximation:] In FoPID, while $\lambda$ provides greater flexibility in design of integrator, the flexibility provided by $\alpha$ in design of derivative action is more useful since this directly relates to robustness and stability of system. The derivative action is bounded to the frequency range $[\omega_d,\omega_t]$. Crone approximation is popular among control engineers since it allows for upper and lower limits to be defined easily and accurately and hence easy approximation of derivative action. Due to its popularity in literature, this is the default approximation technique of FLOreS. Crone approximation is based on a recursive distribution of poles and zeros with the approximation given as:

\begin{equation}
I^\nu(s) = C_0\prod_{k=1}^{N}\frac{1 + \frac{s}{\omega_k^\prime}}{1 + \frac{s}{\omega_k}}
\end{equation}

The values of $\omega_k^\prime$ and $\omega_k$ determine both the frequency range of approximation and order $\lambda$ of integral action. The complete calculation for obtaining these values can be found in \citep{oustaloup2000crone}.

\item[Carlson approximation:] This method presented in \cite{carlson1964approximation} is based on Newton's iterative method and is very accurate. However, its use in practice is limited since it only works when the order $\nu$ is the inverse of an integer such as 1/2 or 1/5. Other fractional orders can be approximated by adding multiple approximations. However, since this method generally leads to higher order transfer functions depending on number of iterations used, the summation of multiple approximations results in very large functions which might be extremely computationally expensive. If we define the fractional operator as $$F(s) = s^\nu$$ then it is approximated iteratively as

\begin{equation}
{F}_{i} (s) = {F}_{i-1} (s) \dfrac{\left(\dfrac{1}{v}-1\right) {F}_{i-1}^{1/v} (s) + \left(\dfrac{1}{v}+1\right)s } { \left( \dfrac{1}{v} + 1 \right) {F}_{i-1}^{1/v} (s) + \left( \dfrac{1}{v} - 1 \right)s}
\end{equation} with $F_0(s) = 1$

\item[Matsuda approximation:] Matsuda approximation is based on approximation of an irrational function by a rational one. Assuming that the logarithmically spaced points are $\omega_k$, $k = 0,1,2....$, approximation takes the form:

\begin{equation}
F(s) = a_0 + \frac{s - s_0}{a_1 + }\frac{s - s_1}{a_2 + }\frac{s - s_2}{a_3 + }....
\end{equation}
where $a_i = v_i(s_i)$, $v_0(s) = s$ and $v_{i+1}(s) = \frac{s - s_i}{v_i(s) - a_i}$

For $s^\nu$ when $|0<\nu<1|$ it is moderately accurate with a limited number of terms and is a good alternative for Crone approximation. 

The bode plots of approximated transfer functions obtained from the 3 methods are plotted in Fig. \ref{fig:cont}. In the case of Crone, since the frequency limits can be set within the method, this is done so. However, in the other 2 cases, the limits can be imposed by choosing the number of iterations. However, this is not accurate as shown. This is true even though Carlson and Matsuda both result in equally or more computationally expensive transfer functions. In the design of both IoPID and FoPID using loop-shaping method, the exact frequency range in which derivative and integral actions perform is important. From the results seen, it is clear that this accuracy is better achieved with Crone. Hence, Crone is the default approximation method of FLOreS.

\begin{figure}
\begin{center}
\includegraphics[width=0.9\columnwidth]{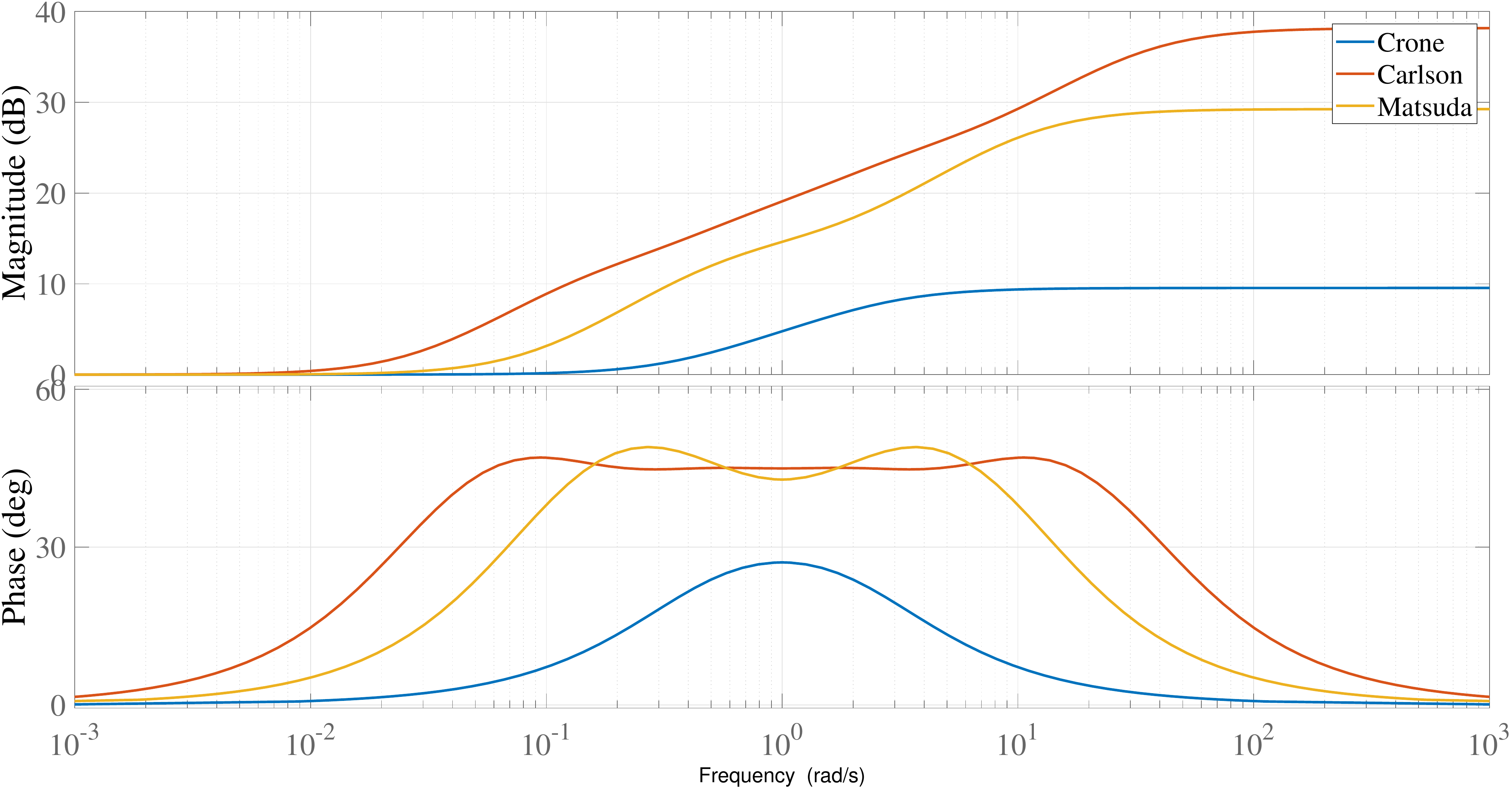}
\caption{Bode plot of fractional order $s^{1/2}$ approximated with the Crone, Carlson and Matsuda method. They have 2, 6 and 2 poles respectively}
\label{fig:cont}
\end{center}
\end{figure}

\item[Discrete approximation methods]

Three discrete approximation methods namely Tustin, Second-order backwards finite difference (SOBFD) and Third-order backwards finite difference (TOBFD) are included in FLOreS. The methods are based on the same concept and consist of 4 steps. \citep{de2005fractional}

\begin{enumerate}
	\item Choosing an equation for conversion from continuous s domain to discrete z domain
	\item Raising it to the desired fractional power 
	\item Expanding the result into a continued fraction
	\item Truncating the series after a reasonable number of terms
\end{enumerate}
The difference between the methods is in the first step over choice of s to z conversion.

\item[Tustin approximation:] Tustin approximation is popular and results in a function which oscillates around the desired values. The equation is given as:

\begin{equation}
	s \approx \dfrac{2}{T} \dfrac{1-z^{-1}}{1+z^{-1}}
	\label{Tus1}
\end{equation}
Applying the next 3 steps gives:

\begin{equation}
\begin{aligned}
&\hat{F}(z^{-1}) = \left(\frac{2}{T}\right)^{v} \Gamma(v+1)\Gamma(-v+1)x \sum_{k=0}^{N} z^{-k}\\
&\left[\sum_{j=0}^{k}\dfrac{(-1)^{j}}{\Gamma(v-j+1)\Gamma(k-j+1)\Gamma(-v+j-k+1)}\right]
\end{aligned}
\label{Tus2}
\end{equation}

\item[Second/Third order backwards finite difference] \item[approximation:] 

The second order backwards finite difference and third order backwards finite difference are common interpolation methods.

Second order function is given as:
\begin{equation}
	s \approx \dfrac{3-4z^{-1}+z^{-2}}{2T}
	\label{SO1}
\end{equation}
Applying the next 3 steps gives:

\begin{equation}
	\begin{aligned}
		&\hat{F}(z^{-1}) = \frac{[\Gamma(v+1)]^{2}}{(2T)^{v}}x 
		\sum_{k=0}^{N} z^{-k} \sum_{j=0}^{k}\\
		&\dfrac{3^{v-j}(-1)^{k}}{\Gamma(j+1)\Gamma(v-j+1)\Gamma (k-j+1)\Gamma(v-k+j+1)}
	\end{aligned}
	\label{SO2}
\end{equation}

While TOBFD is slightly more accurate than SOBFD, it sometimes also generates transfer functions with complex parts which are not usable. The third order function is given as:

\begin{equation}
	s \approx \dfrac{11-18z^{-1}+9z^{-2}-2z^{-3}}{6T}
	\label{TO1}
\end{equation}
Applying the next 3 steps gives:
\begin{equation}
	\begin{aligned}
		&\hat{F}(z^{-1}) = \frac{[\Gamma(v+1)]^{3}}{(3T)^{v}} \sum_{k=0}^{N} z^{-k}
		\bigg[g[\sum_{t=0}^{k} \sum_{p=0}^t\\  
		&\dfrac{(-1)^{p}}{\Gamma(p+1)\Gamma(v-p+1)\Gamma(t-p+1)}\\
		&\dfrac{(-\dfrac{7}{4}-\dfrac{\sqrt{39}}{4}j)^{v-t+p}
			((-\dfrac{7}{4}-\dfrac{\sqrt{39}}{4}j)^{v-k+t})}
		{\Gamma(v-t+p+1)\Gamma (k-t+1)\Gamma(v-k+t+1)}\Bigg]
	\end{aligned}
	\label{TO2}
\end{equation}

\end{description}

\begin{figure}
\begin{center}
\includegraphics[width=0.9\columnwidth]{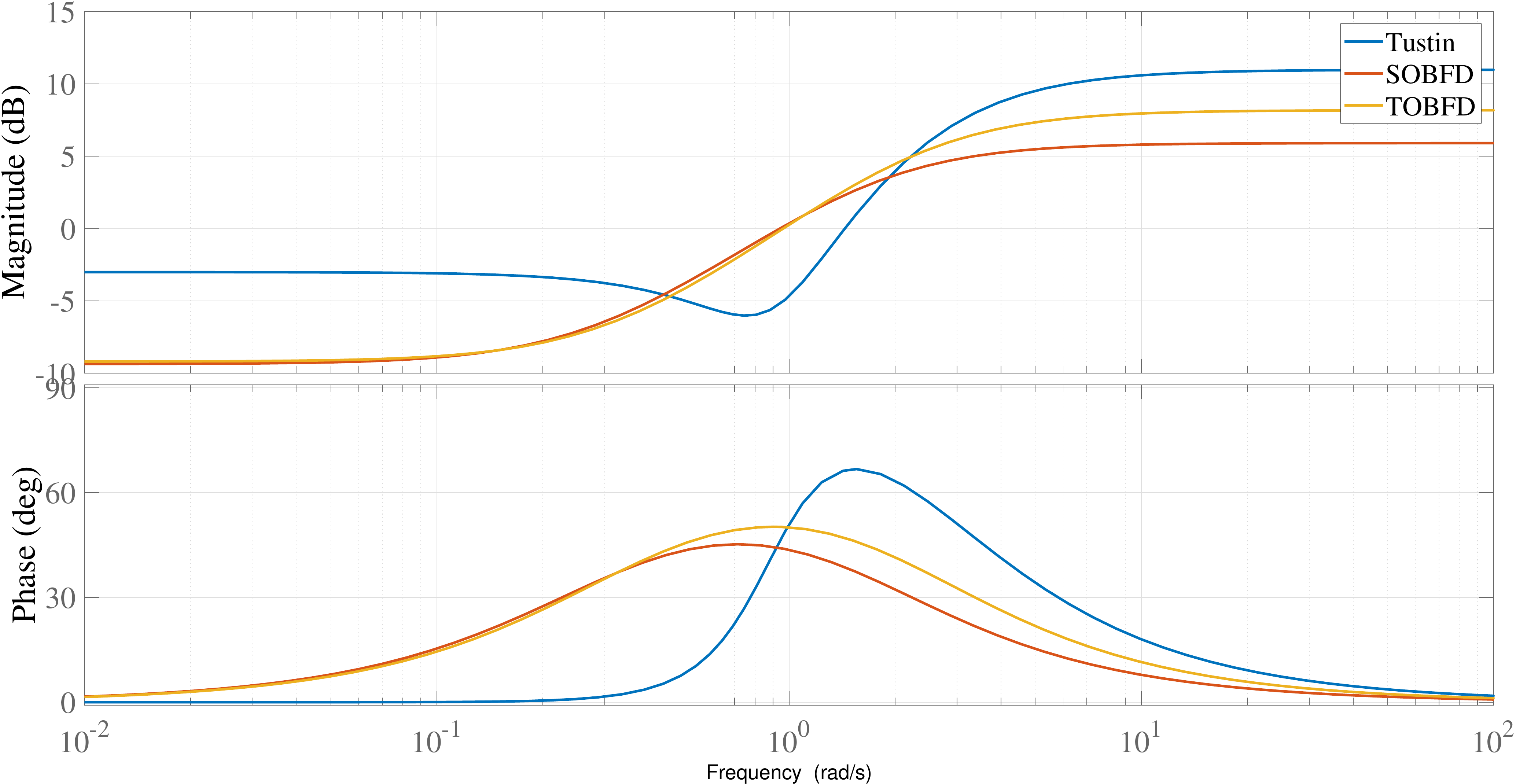}
\caption{Bode plost of fractional order $s^{1/2}$ approximated with the Tustin, SOBDF and TOBDF. They all have 3 poles.}
\label{fig:dis}
\end{center}
\end{figure}

The approximation obtained from these 3 methods is shown in Fig. \ref{fig:dis}. 

\section{Loop-shaping Toolbox --- FLOreS}

\begin{figure}
	\begin{center}
		\includegraphics[width=\linewidth]{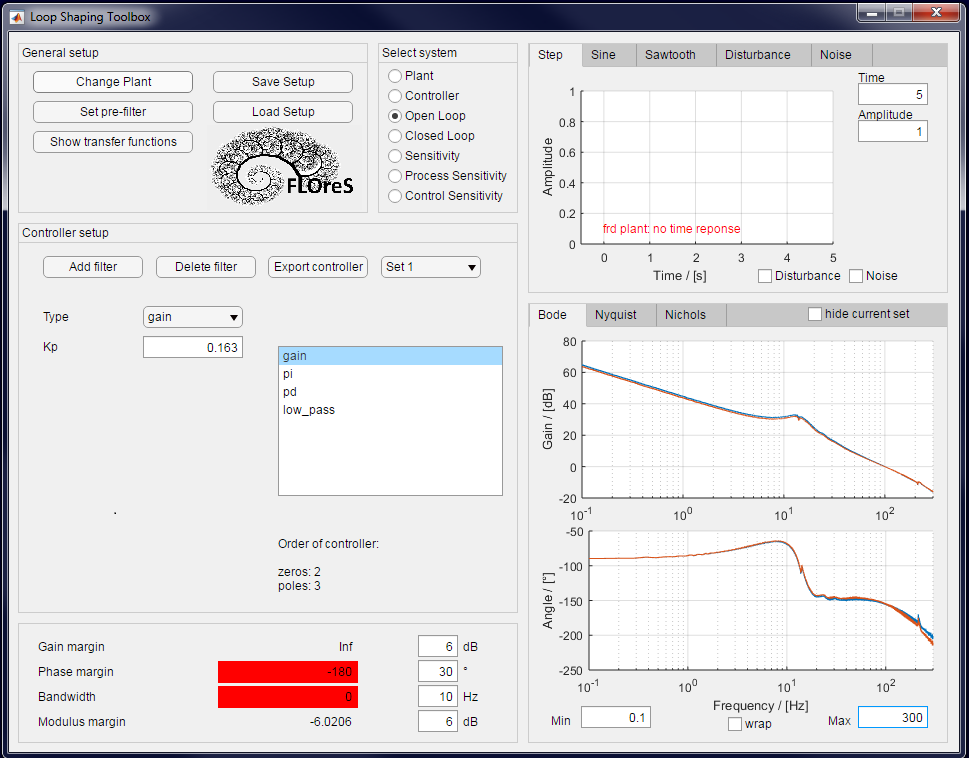}    
		\caption{Screenshot of FLOreS while tuning controllers for a $4^{th}$ order system} 
		\label{fig:gui1}
	\end{center}
\end{figure}

FLOreS has been developed to enable loop-shaping for fractional order controllers for both motion and process control. The screenshot of the user interface is shown in Fig. \ref{fig:gui1}.The features of FLOreS are described below.

\begin{description}[leftmargin=*]
\item[Plant] 
A SISO system or plant can be imported to the toolbox as transfer function from workspace. However, in the industry, plant frequency response data obtained directly is often used for tuning. Hence, FLOreS is designed to import frd directly. Since FLOreS has also been designed to be used as an educational tool, example plants like mass-spring-damper systems are included and can be selected. In this scenario, the variable values can be set and the transfer function is generated and imported automatically.

\item[Controller] 
The most important feature of FLOreS is in the availability of fractional order filters for controller design. While integer order filters like \textit{pd}, \textit{pi}, \textit{pid}, \textit{lead lag}, \textit{notch} are made available, fractional variants of all these filters can also be used in FLOreS. Additionally, integer order low pass filters can also be used. As noted earlier, all fractional variants make use of crone approximation. Additionally, a pre-filter can also be imported from workspace. 

Apart from the fractional variants of filters made available, all the 6 approximation methods can also be used separately to design separate filters.

The order of designed controllers is displayed to provide an indication of computational requirements especially since approximation can lead to higher order controllers.

Multiple controllers can be designed within FLOreS simultaneously allowing for comparison of different performance aspects.

\item[Frequency response] 
The frequency responses can be seen as either \textit{Bode} (with wrapping of phase being an option), \textit{Nyquist} or \textit{Nichols} plots.The plots of one of the following subsystems can be visualized at any time. They are \textit{Plant}, \textit{Controller}, \textit{Open~Loop}, \textit{Closed~Loop}, \textit{Sensitivity}, \textit{Process~Sensitivity} and \textit{Control~Sensitivity}. These plots play a crucial role in loop-shaping since they allow for an intuitive graphical approach to designing controllers. Further the frequency domain approach allows for multiple performance aspects like stability, disturbance rejection and noise attenuation to be analysed within the same tool. 

The visible frequency range is made adjustable. In case the FRD of plant is imported, then this range is set automatically for the available data. 

\item[Performance] 
In the performance panel \textit{gain~margin}, \textit{phase~margin}, \textit{modulus~margin} and \textit{system~bandwidth} are displayed. Controller requirements can be defined by the engineer and corresponding performance values will be highlighted when the requirements aren't met.

\item[Time response] 
The time response of closed loop system to \textit{step}, \textit{sine} or \textit{sawtooth} reference signal is plotted in this window.

Additionally the response of system to \textit{step}, \textit{sine} or \textit{gaussian} disturbance signal or a \textit{sine} or \textit{gaussian} noise signal can also be visualized. The time response of system for either of the references in combination with selected disturbance and/or noise can also be seen. This feature is novel to FLOreS and helps the engineer get a feeling for the system response for real-world conditions.

If a pre-filter is defined, the time response with and without the pre-filter will be displayed to allow for comparison and performance assessment.

However, this function is not available when the plant is loaded using frequency response data.

\item[Additional features] 
The transfer function of the controller can be exported to the MATLAB workspace allowing for easy implementation. Further, the complete design session on FLOreS can be saved as a '.lstb' file allowing for restarting the session at a different time or for engineers to work in groups.

\end{description}

\section{FLOreS - Example}

\begin{figure}[H]
	\begin{center}
		\includegraphics[width=\linewidth]{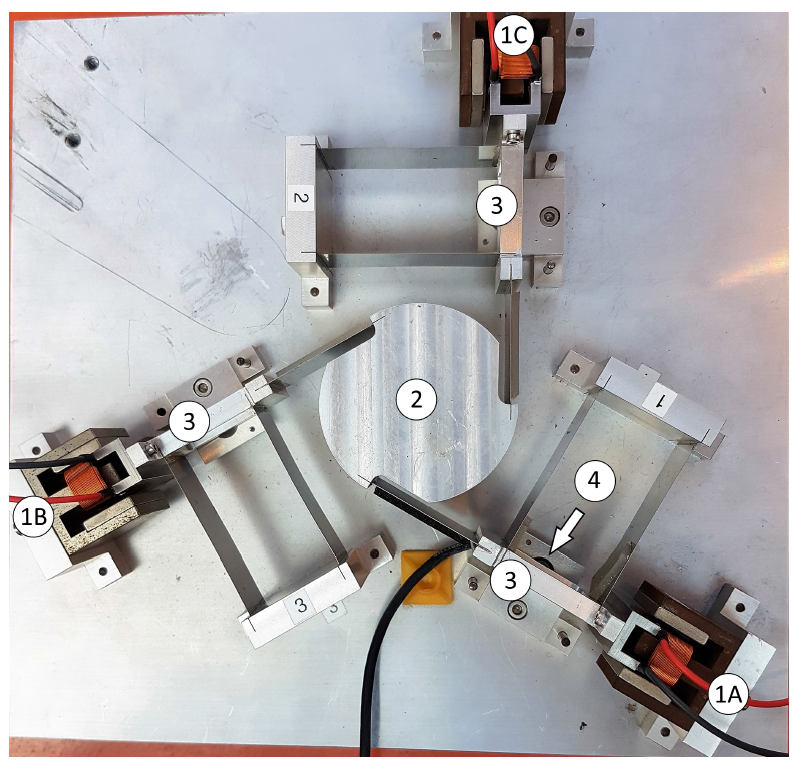}
		\caption{3 DOF planar precision positioning stage used for testing FLOreS} 
		\label{Spyder}
	\end{center}
\end{figure}

FLOreS is tested for designing controllers for a precision positioning stage called 'Spyder Stage' shown in Fig. \ref{Spyder}. The stage is actuated using Lorentz coil actuators shown as 1A, 1B and 1C. 3 masses indicated by number 3 in figure are connected to the real world using leaf flexures providing large stiffness in 5 DOFs and relatively low stiffness in the required DOF. Encoders placed below each of these masses allow for accurate position measurement. All the 3 masses are connected to central mass indicated by number 2 with one leaf flexure each. The design of stage allows 3 DOF planar movement of mass 2. Although this stage allows for 3 DOF planar positioning, only one of the DOFs is considered and used for the purpose of testing FLOreS. Actuator 1A is selected for this purpose.

\begin{figure}
\begin{center}
\includegraphics[trim = {2.5cm, 0cm, 2.5cm, 0cm},width=\linewidth]{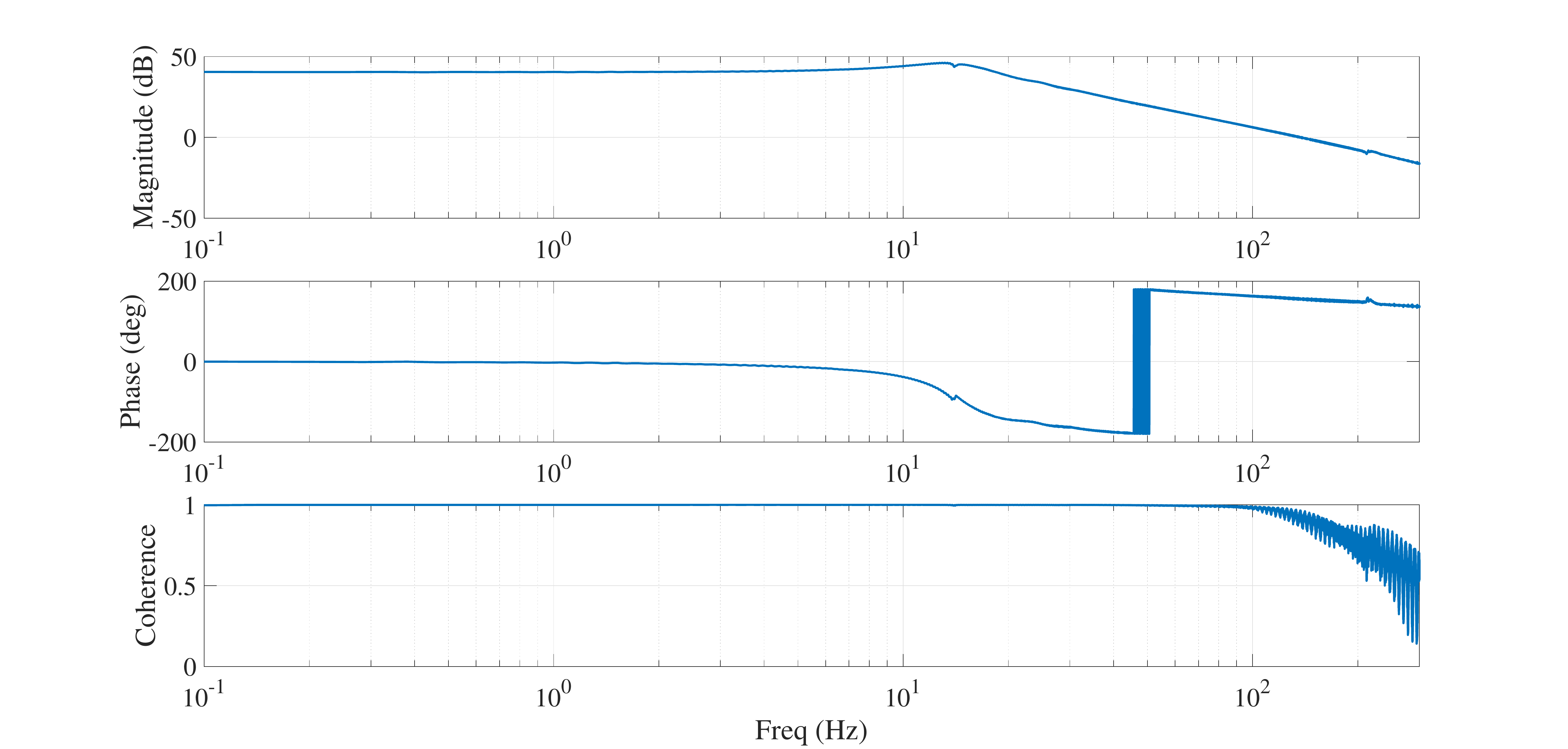}
\caption{Frequency response data of the precision stage.} 
\label{FRD}
\end{center}
\end{figure}

Since modelling this system can be cumbersome, frequency response of plant is obtained by applying a Chirp signal. The FRD of system is shown in Fig. \ref{FRD}. The system has the behaviour of a collocated double mass-spring system. Although FLOreS is created to enable design of FOCs, an IOC and an FOC are designed for this system for testing. The parameters of both controllers are given in Table. \ref{IntegerTable} and Table. \ref{FractionalTable} respectively. In the case of FOC, a fractional order derivative action is employed along with fractional order filtering in range $[1000, 10000\ Hz]$ using a fractional lead-lag filter. Since the controllers are designed for testing FLOreS, both FOC and IOC have similar behaviour with FOC having slightly higher phase margin and higher gain at low frequencies.

{\renewcommand{\arraystretch}{1.5}%
\begin{table}
\begin{center}
\caption{Parameters of the integer order controller}
\label{IntegerTable}
\begin{tabular}{l|llll|}
Type		&Parameters							\\ \hline
gain 		&Kp: 0.163 			& 				\\ \hline
pi 			&$f_i$:  10			&	  			\\ \hline
pd 			&$f_d$: 33.33 		&$f_t$: 300 	\\ \hline
low pass 	&$f_{cutoff}$: 1000 	&order: 1 		\\ 
\end{tabular}
\end{center}
\end{table}

\begin{table}
\begin{center}
\caption{Parameters of the fractional order controller}
\label{FractionalTable}
\begin{tabular}{l|llll}
Type			&Parameters													\\ \hline
gain 			&Kp: 0.0023													\\ \hline
pi 				&$f_i$: 10 			&  				& 				& 		\\ \hline
frac. pd 		&$f_d$: 33.33		&$f_t$: 300		&$\alpha$: 1.1 	&N: 3	\\ \hline
frac. leadlag	&$f_z$: 10000 		&$f_p$: 1000 	&$\alpha$: 1.8  	&N: 3   \\ \hline
low pass		&$f_{cutoff}$: 10000	&order: 1		&				&	  	\\ 
\end{tabular}
\end{center}
\end{table}

The obtained controllers are implemented on NI Myrio modules which allow for real-time implementation on FPGA. The step responses obtained for both controllers are shown in Fig. \ref{fig:Step}. As noted earlier, since the controllers have similar behaviour, this is also seen in the step response. The slightly higher gain of FOC translates to better settling than IOC. Chirp reference signals are applied to the system in closed-loop to obtain closed-loop complementary sensitivity functions. These are compared against the estimated ones shown in FLOreS in Fig. \ref{fig:ClosedLoop}. The estimated and measured frequency response matches very well as seen in the figure and shows that FLOreS can be used to advance FOCs in the industry.

\begin{figure}
	\begin{center}
		\includegraphics[width=\linewidth]{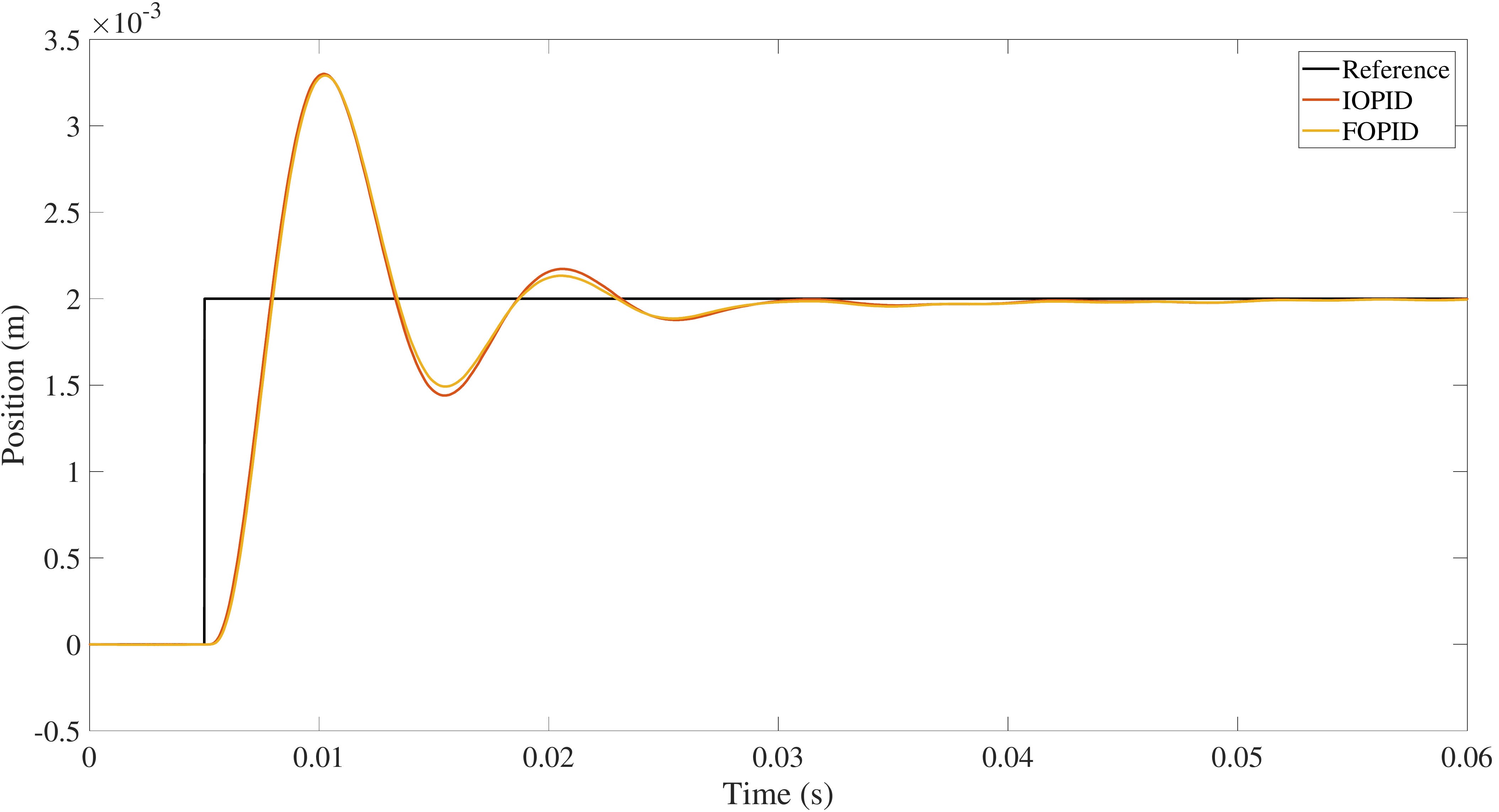}
		\caption{Step response of both the IOC and FOC controlled system.}
		\label{fig:Step}
	\end{center}
\end{figure}

\begin{figure}
\begin{center}
\includegraphics[trim = {2.5cm, 0cm, 2.5cm, 0cm},width=\linewidth]{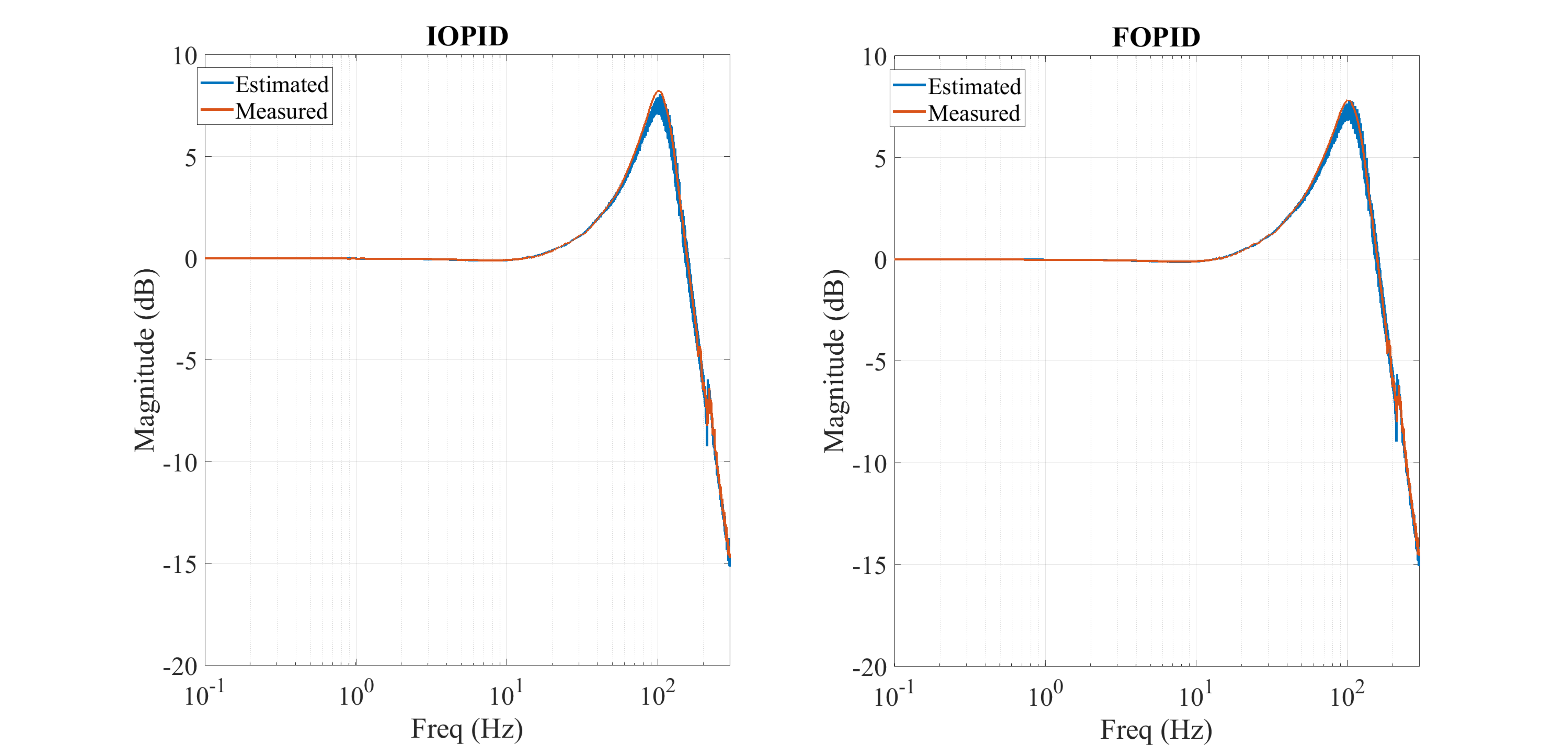}
\caption{Closed loop bode plot of both the integer order and fractional order controlled system.} 
\label{fig:ClosedLoop}
\end{center}
\end{figure}

\section{Conclusion}

FLOreS is the first toolbox designed to allow designing fractional order controllers using the industry standard loop shaping method. This allows for intuitive graphical approach to design and assessment of different performance aspects. FLOreS is designed to allow frequency response data of practical plant to be directly used during design. The main novelty of FLOreS is the availability of fractional order filters which use crone as the default approximation method. Apart from this, 5 other approximation methods can also be used.

Open loop and closed loop sensitivity functions are displayed graphically to assist the control engineer. Additionally, stability margins and bandwidth are also displayed. In the case where plant is imported as a transfer function, time domain response of closed-loop plant to references, noise and disturbances can also be visualized within FLOreS.

The toolbox is tested to design both IOC and FOC for one of the DOFs of a planar precision positioning stage. The FRD obtained practically is used for this purpose. The designed controllers are tested on the setup and closed loop frequency response obtained in practice is compared with the one estimated by FLOreS. It is seen that the results match well, showing industry readiness of toolbox.

The toolbox is available for use under fair usage policy on the department website. Please contact the corresponding author (****) for more information.

\bibliography{research}

\end{document}